\documentclass[10pt,a4paper,twocolumn,showpacs]{revtex4}
\usepackage{graphicx}
\usepackage{amsfonts}
\usepackage{amsbsy}
\newcommand{\beq}{\begin{equation}}
\newcommand{\eeq}[1]{\label{#1}\end{equation}}

\newcommand{\be}{\begin{eqnarray}}
\newcommand{\ee}{\end{eqnarray}}

\begin{document}

\title{Evidence for the two pole structure of the 
$\Lambda(1405)$ resonance
}

\author{
V.K. Magas$^1$, E. Oset$^1$, A. Ramos$^2$\\}

\affiliation{
$^1$ Departamento de F\'{\i}sica Te\'orica and IFIC, Centro Mixto,\\ 
     Institutos de Investigaci\'on de Paterna - Universidad de Valencia-CSIC\\ 
     Apdo. correos 22085, 46071, Valencia, Spain\\
$^2$ Departament d'Estructura i 
Constituents de la Mat\'eria, \\
Universitat de Barcelona,
 Diagonal 647, 08028 Barcelona, Spain
}

\begin{abstract}
The $K^- p \to \pi^0 \pi^0 \Sigma^0$ reaction is studied within a chiral
unitary model. The distribution of $\pi^0 \Sigma^0$
states forming the $\Lambda(1405)$ shows, in agreement with a recent experiment,
a peak at $1420$ MeV and a relatively
narrow width of $\Gamma = 38$ MeV. 
The mechanism for the reaction is largely
dominated by the emission of a $\pi^0$ prior to the $K^- p$ interaction
leading to the $\Lambda(1405)$. This ensures the coupling of the  
$\Lambda(1405)$ to the $K^- p$ channel, thus maximizing the contribution of the
second state found in chiral unitary theories, which is narrow and of higher 
energy than the nominal $\Lambda(1405)$. This is unlike the $\pi^- p \to K^0 \pi
\Sigma$ reaction, which gives more weight to the pole at lower energy and with a
larger width. The data of these two experiments, together with the present theoretical
analysis, provides a firm evidence of the two pole structure of the 
$\Lambda(1405)$.
\end{abstract}


\maketitle

The history of the $\Lambda(1405)$ as a dynamical resonance generated
from the interaction of meson baryon components in coupled channels is
long \cite{Jones:1977yk}, but it has experienced a boost within the context
of unitary extensions of chiral perturbation theory ($U\chi PT$) 
\cite{Kaiser:1995cy,Kaiser:1996js,kaon,Oller:2000fj,Jido:2002yz,Garcia-Recio:2002td}, where
the  lowest order chiral Lagrangian and unitarity in
coupled channels generates the $\Lambda(1405)$  and leads to good
agreement with the $K^- p$ reactions.  The surprise, however, came with the
realization that there are two poles in the neighborhood of the 
$\Lambda(1405)$ both contributing to the final experimental invariant mass
distribution \cite{Oller:2000fj,Jido:2002yz,Garcia-Recio:2002td,
Jido:2003cb,Garcia-Recio:2003ks,Hyodo:2002pk,Nam:2003ch}.
The properties of these two states are quite different, one has a mass around 
$1390$ MeV, a large
width of about $130$ MeV and couples mostly to $\pi \Sigma$, while the second
one has a mass around $1425$ MeV, a narrow width of about $30$ MeV and couples
mostly to $\bar{K} N$. The  two states are populated
with different weights in different reactions and, hence, their superposition can lead to
different distribution shapes. Since the $\Lambda(1405)$ resonance is always seen from the
invariant mass of its only strong decay channel, the $\pi \Sigma$,
hopes to see the second pole are tied to having a reaction where
the $\Lambda(1405)$ is formed from the $\bar{K} N$ channel.  In this sense a
calculation of the $K^- p \to \gamma \Lambda(1405)$ reaction \cite{Nacher:1999ni}, prior to the 
knowledge of the existence of the two $\Lambda(1405)$ poles,
showed a narrow structure at about 1420
MeV.  With the present perspective this is clearly interpreted as
the reaction proceeding through the emission of the photon followed by 
the generation of the resonance  from $K^- p$, thus receiving a large
contribution from the second narrower state at higher energy. 
The same idea is used in the reaction
$\gamma p \to K^* \Lambda(1405)$ \cite{Hyodo:2004vt} which proceeds via K-meson
exchange and is now under
investigation at Spring8/Osaka \cite{nakano}. Luckily, the recently measured reaction
$K^- p \to \pi^0 \pi^0 \Sigma^0$  \cite{Prakhov} allows us to test already the
two-pole nature of the $\Lambda(1405)$. This
process
shows a strong similarity with the reaction $K^- p \to \gamma \Lambda(1405)$, where the
photon is replaced by a $\pi^0$. 

\begin{center}
\begin{figure}[htb]
\vspace{-0.7cm}
    \includegraphics[height=3.5cm,width = 5.5cm]{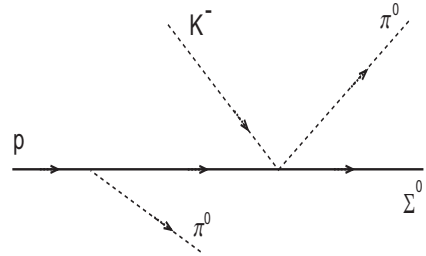}
\vspace{-0.3cm}
 \caption{\label{fig:tree}
Nucleon pole term for the $K^- p \to \pi^0 \pi^0 \Sigma$ reaction.}
\vspace{-1.1cm}
\end{figure}
\end{center}

\begin{center}
\begin{figure}[htb]
    \includegraphics[width = 8.4cm]{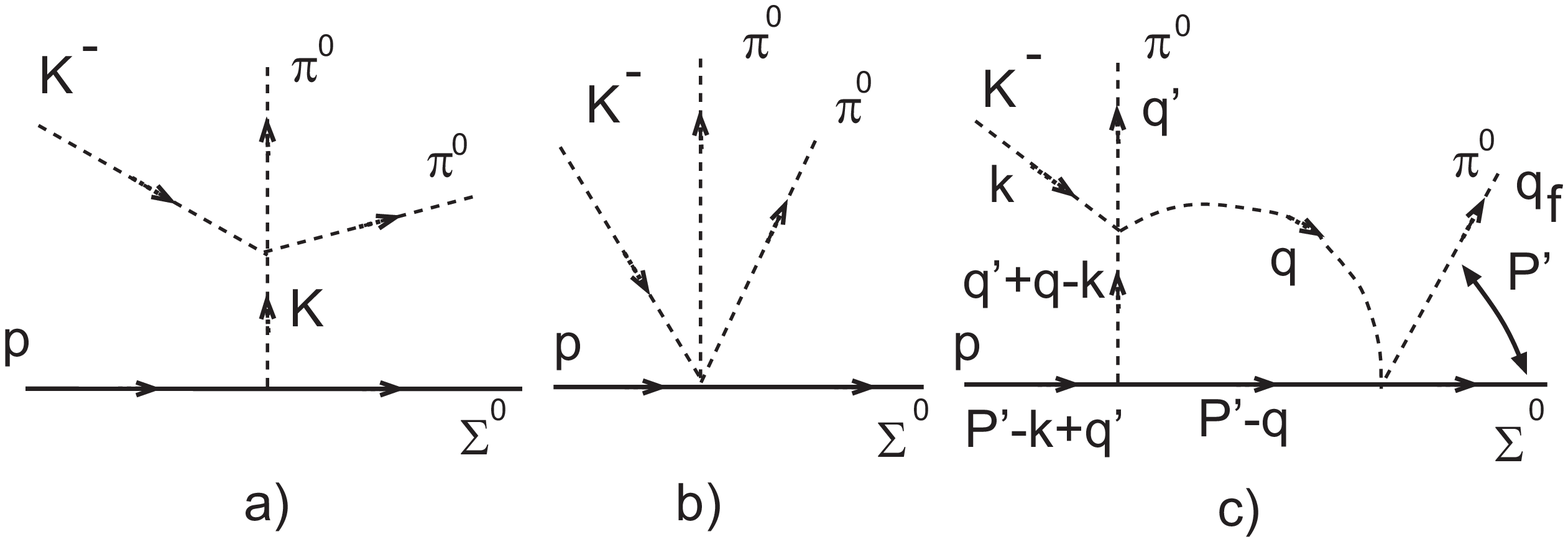}
\vspace{-0.4cm}
 \caption{\label{fig:loop}
Diagrams for the $K^- p \to \pi^0 \pi^0 \Sigma$ reaction: a) K-pole term; b) MMMBB diagram - canceled by the off-shell part of a) diagram;
c) one loop contribution. }
\end{figure}
\end{center}

Our model for the reaction 
$K^- p \to \pi^0 \pi^0 \Sigma^0 $
in the energy region of $p_{K^-}=514$ to $750$ MeV/c, as in the experiment \cite{Prakhov}, 
considers those mechanisms in which a $\pi^0$ loses the necessary energy
to allow the remaining $\pi^0\Sigma^0$ pair to be on top of the $\Lambda(1405)$
resonance.  The first of such mechanisms is given by the diagram of
Fig.~\ref{fig:tree}.
 In addition, analogy with the $\pi^- p \to K^0 \pi \Sigma$ reaction, where
the $\pi\Sigma$ is also produced in the $\Lambda(1405)$ region, demands that one
also considers the mechanisms of Fig.~\ref{fig:loop}.
Indeed, the analogous diagrams replacing $(K^-,\pi^0)$ by
$(\pi^-,K^0)$ were evaluated in Ref.~\cite {hyodo} giving
a sizable 
contribution to the cross section. One of the technical findings of
Ref.~\cite{hyodo} is that the diagram of Fig.~\ref{fig:loop}b canceled exactly
the off-shell part of the meson meson amplitude in Fig~\ref{fig:loop}a, which
therefore requires to be evaluated with the meson meson amplitude writen on
shell. We should also note that the equivalent diagram of Fig.~\ref{fig:tree} in
the case of the $\pi^- p \to K^0 \pi \Sigma$ reaction is strongly reduced since
the emission of a $K^0$ requires an intermediate hyperon ($ \Lambda$,$\Sigma$)
state very far off shell. This is not the case in the present reaction where the
intermediate proton is only moderately off shell.

The amplitude for the Feynman diagram of Fig.~\ref{fig:loop}c is given by
$$
-i t =\int \frac{d^4q}{(2\pi)^4} (-i t_{K^- m \to \pi^0 {m^\prime}}) \times
$$
$$
 \frac{i}{(q+q^\prime-k)^2 - M^2_m+i\varepsilon}
\frac{i}{q^2 - M^2_{m^\prime}+i\varepsilon} 
\frac{M_B}{E_B(q)} \times
$$
\beq
\frac{i}{P^{\prime\, 0} -q^0-E_B(q) +i\varepsilon} 
 {\cal C}_{p B m^\prime} \vec{\sigma} (\vec{q}\,^\prime + \vec{q}-\vec{k} ) (-i t_{m^\prime B \to
\pi^0\Sigma^0}) \ ,
\eeq{eq:loop}
where $t_{K^- m \to \pi^0 m^\prime}$ refers to the meson meson amplitude, 
which we take from
the lowest order chiral lagrangian \cite{gasser}, the
$t_{m^\prime B \to \pi^0\Sigma^0}$ are the meson baryon amplitudes taken from
the model of Ref.~\cite{kaon}, and 
the coefficients ${\cal C}_{p
B m^\prime}$ are obtained from the $D,F$ terms of the meson baryon lagrangian
\cite{ulf,ecker} at lowest order. We take $D+F=1.26$ and $D-F=0.33$.
There are the
following eight meson meson baryon $(m m^\prime B)$
intermediate states: $K^+\pi^0\Sigma^0$, $K^+\pi^0\Lambda$, $\pi^0 K^- p$,
$ K^0 \pi^- \Sigma^+$, $\pi^+ {\bar
K}^0 n$, $K^+ \eta\Sigma^0$, $K^+ \eta\Lambda$, $\eta K^- p$. 
We note that the $t_{K^- m \to \pi^0 m^\prime}$ amplitude 
is largely dominated by the s-wave and that the linear terms in $\vec{q}$ are further reduced 
in the loop due to the s-wave character of the  $t_{m^\prime B \to \pi^0\Sigma^0}$  amplitude. 
Hence, one can factorize 
the on shell $t_{K^- m \to \pi^0 m^\prime}$ amplitude outside the integral in
Eq.~(\ref{eq:loop}) and perform the $q^0$ integration analytically. 
The $t_{m^\prime B \to \pi^0\Sigma^0}$  amplitude depends on the ($\pi^0 \Sigma^0$) invariant mass, $M_I^2=(q_f+p_\Sigma)^2$, 
and also factorizes outside the integral.

\begin{center}
\begin{figure*}[htb]
\vspace{-0.5cm}
   \includegraphics[width = 16.8cm]{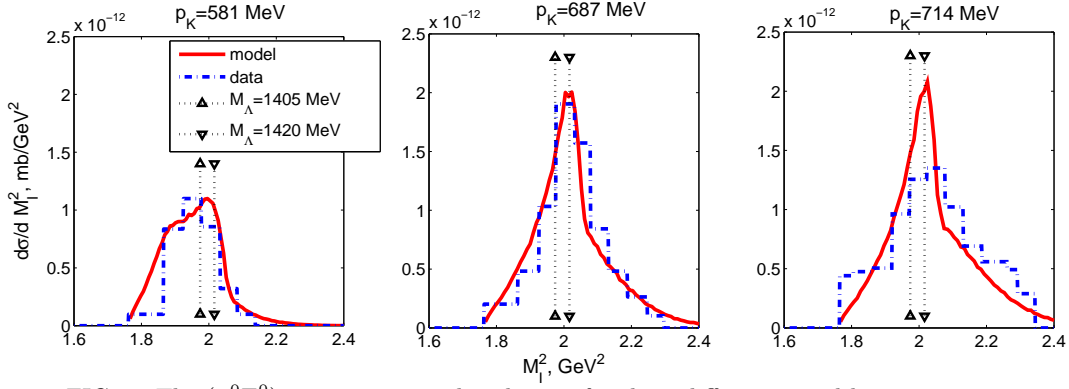}
 \vspace{-0.5cm}
 \caption{The ($\pi^0 \Sigma^0$) invariant mass distribution for three different initial kaon
momenta.
 \label{fig:mass}
 }
\end{figure*}
\end{center}


Altogether the amplitude for the process $K^- p \to \pi^0 \pi^0 \Sigma^0 $, considering
the diagram of Fig.~\ref{fig:tree} with a pseudovector $\pi N N$ coupling (and keeping 
up to $1/M_N$ terms) and those of Figs.~\ref{fig:loop}a and
\ref{fig:loop}c, is given by:
\begin{equation}
-i t_{K^- p \to \pi^0 \pi^0 \Sigma^0 } = -i t^{(N-{\rm pole})}- i t^{(K-{\rm
pole})}-i
t^{({\rm loop})} \ ,
\label{eq:ampl}
\end{equation}
with
\begin{eqnarray}
&&- i t^{(N-{\rm pole})}= -  \frac{D+F}{2f} 
\vec{\sigma}\left[\vec{q}\,^\prime (1+\frac{q^0\,^\prime}{2 M_N}) + \frac{q^0\,^\prime}{M_N} \vec{k}\right]
\times
\nonumber \\
&& 
\frac{M_N}{E_N(\vec{k}+\vec{q}\,^\prime)}\frac{1}{ E_N(\vec{k})-q^{\prime\, 0} -
E_N(\vec{k}+\vec{q}\,^\prime)}
t_{K^- p \to \pi^0\Sigma^0}  ,
\label{eq:npole}
\end{eqnarray}

\begin{eqnarray}
-i t^{(K-{\rm pole})} &=&  - \frac{1}{4f^2}(2 M_\pi^2 + 2 q^\prime q_f)
\frac{1}{(k-q^\prime-q_f)^2-M_K^2}\times
\nonumber \\
&&  ~~~\frac{D-F}{2f}
\vec{\sigma}(\vec{k}-\vec{q}\,^\prime-\vec{q}_f) \ ,
\label{eq:kpole}
\end{eqnarray}
\begin{eqnarray}
-i t^{({\rm loop})} &=& \sum_{j=1}^{8} t_{K^- m \to \pi^0 m^\prime}^{(j)} {\cal
C}^{(j)}_{p B m^\prime}
\vec{\sigma}(\vec{k}-\vec{q}\,^\prime) t_{m^\prime B \to
\pi^0\Sigma^0}^{(j)} \nonumber \\
&& ~~~~~~~\times
I(k,q^\prime,M_I,M^j_m,M_{m^\prime}^j,M_B^j)\ ,
\label{eq:loop2}
\end{eqnarray}
where 
$I(k,q^\prime,M_I,M^j_m,M_{m^\prime}^j,M_B^j)$ is the result of the $q^0$ integral in
Eq.~(\ref{eq:loop}), which in the reference frame where $\vec{P}^\prime=0$ 
(see Fig. \ref{fig:loop}c) is given by:
\begin{widetext}
\begin{eqnarray}
I(k,q^\prime,M_I,M_m,M_{m^\prime},M_B)=
\int \frac{d^3q}{(2\pi)^3} \frac{1}{2\omega\omega^\prime}
\frac{M_B}{E_B(q)}\left[(\omega+\omega^\prime)^2+(\omega+\omega^\prime)
(E_B(q)-P^{\prime\,0})+\omega(k^0-q^{\prime\,0})
\right]
 \left[1 + \frac{\vec{q}\,(\vec{q}\,^\prime-
\vec{k})}{(\vec{q}\,^\prime-\vec{k}\,)^2} \right]
\nonumber \\
 \times \frac{1}{P^{\prime\,0}-k^0+q^{\prime\,0}-\omega-E_B(q)+i\varepsilon}
\frac{1}{P^{\prime\,0}-\omega^\prime-E_B(q)+i\varepsilon} 
\frac{1}{\omega^\prime+q^{\prime\,0}-k^0+\omega+i\varepsilon}
\frac{1}{k^0-q^{\prime\,0}+\omega+\omega^\prime-i\varepsilon} \, ,
\label{eq:integral}
\end{eqnarray}
\end{widetext}
with $\omega = \sqrt{(M_m)^2+(\vec{q}+\vec{q}\,^\prime-\vec{k}\,)^2} $ and
$\omega^\prime=\sqrt{(M_{m^\prime})^2+\vec{q}\,^2}$. The coefficients ${\cal
C}_{p B m^\prime}$ and the meson meson amplitudes  $t_{K^- m \to \pi^0
m^\prime}$ for the allowed $(m m^\prime B)$ channels are displayed in
Table~\ref{tab:tab1}.

\begin{center}
\begin{figure}[htb]
\vspace{-0.5cm}
   \includegraphics[width = 6.0cm]{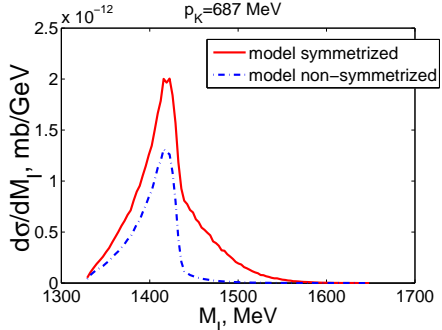}
\vspace{-0.5cm}
 \caption{Theoretical ($\pi^0 \Sigma^0$) invariant mass distribution for an initial kaon lab
 momenta of $687$ MeV. The non-symmetrized distribution also contains the factor 1/2
 in the cross section.
 \label{tal}
 }
\vspace{-0.5cm}
\end{figure}
\end{center}

\begin{table*} \begin{center} \caption{The $t_{K^- m \to \pi^0 m^\prime}$
	amplitudes  and 
	${\cal C}_{p B m^\prime}$ coefficients for the allowed ($m m^\prime B$) channels 
	(see Eq.~\ref{eq:loop2})). Here $q^0_{mB}=\frac{M_I^2+M_m^2-M_B^2}{2M_I}.$ }
	\begin{tabular}{ccc|c|c} $m$  & $m^\prime$  & $B$  & $t$ & ${\cal C}$\\
	\hline $K^+$  & $\pi^0 $  & $\Sigma^0 $  &
	$-\frac{1}{4f^2}\left(2M_\pi^2+2q'^0 q^0_{\pi\Sigma}\right) $  &
	$\frac{D-F}{2f} $ \\ $ K^+$  & $\pi^0 $  & $\Lambda $  &
	$-\frac{1}{4f^2}\left(2M_\pi^2+2q'^0 q^0_{\pi\Lambda}\right) $ &
	$-\frac{2}{\sqrt{3}}\frac{D-F}{2f}+\frac{1}{\sqrt{3}}\frac{D+F}{2f}  $
	\\ $ \pi^0$  & $K^- $  & $p $  & $ -\frac{1}{4f^2}\left(2M_K^2-2k^0
	q^0_{Kp}\right) $  & $ \frac{D+F}{2f} $\\ $ K^{0}$  & $\pi^- $  &
	$\Sigma^+$  & $-\frac{\sqrt{2}}{2f^2}\left(k^0
	q^0_{\pi\Sigma}-k^0q'^0+\vec{k}\vec{q'}\right)  $  &
	$\sqrt{2}\frac{D-F}{2f}  $ \\ $ \pi^+ $  & $\overline{K}^{\ 0} $  & $ n
	$  &
	$-\frac{\sqrt{2}}{2f^2}\left(M_K^2-M_\pi^2-q'^0q^0_{Kn}-k^0q'^0+\vec{k}\vec{q'}
	\right)  $  & $\sqrt{2}\frac{D+F}{2f}$  \\ $K^+ $  & $\eta $  &
	$\Sigma^0 $  & $
	-\frac{\sqrt{3}}{12f^2}\left(3t-\frac{8}{3}M_K^2-\frac{1}{3}M_\pi^2-M_\eta^2
	\right)  $,  & $\frac{D-F}{2f}  $  \\ &  &  & where $
	t=M_\pi^2+M_\eta^2+2q'^0q^0_{\eta\Sigma}$ & \\ $ K^+ $  & $\eta $  &
	$\Lambda$  & $
	-\frac{\sqrt{3}}{12f^2}\left(3t-\frac{8}{3}M_K^2-\frac{1}{3}M_\pi^2-M_\eta^2
	\right)  $,  &
	$-\frac{2}{\sqrt{3}}\frac{D+F}{2f}+\frac{1}{\sqrt{3}}\frac{D-F}{2f}  $
	\\ &  &  & where $ t=M_\pi^2+M_\eta^2+2q'^0q^0_{\eta\Lambda}$ & \\
	$\eta $  & $K^- $  & $ p $  & $
	-\frac{\sqrt{3}}{12f^2}\left(3t-\frac{8}{3}M_K^2-\frac{1}{3}M_\pi^2-M_\eta^2
	\right)  $, & $
	\frac{1}{\sqrt{3}}\frac{D+F}{2f}-\frac{2}{\sqrt{3}}\frac{D-F}{2f} $ \\
	&  &  & where $ t=2M_K^2-2k^0q^0_{Kp}$ &

                \label{tab:tab1}                
                \end{tabular}
        \end{center}

\end{table*}

The indistinguishability of the two emitted pions requires the implementation
of symmetrization. This is achieved by summing two amplitudes evaluated
with the two pion momenta exchanged, $q_f \leftrightarrow q^\prime$, except for
the diagram of Fig.~\ref{fig:loop}a which is already symmetrized. In addition,
a factor of 1/2 for indistinguishable particles is also included in the
total cross section.

Our calculations show that the process is largely dominated by the nucleon
pole term shown in Fig.~\ref{fig:tree}. As a consequence, the $\Lambda(1405)$ thus obtained comes
mainly from the $K^- p \to \pi^0 \Sigma^0$ amplitude which, as mentioned above,
gives  the largest possible weight  to the second (narrower) state.

\begin{center}
\begin{figure}[htb]
\vspace{-0.5cm}
   \includegraphics[width = 6.0cm]{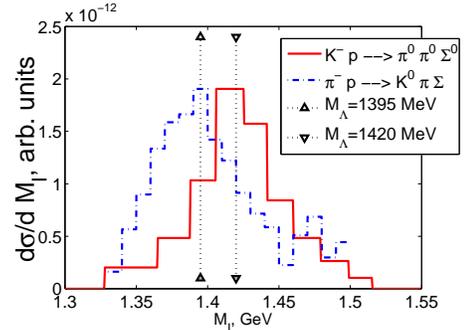}
\vspace{-0.5cm}
 \caption{Two experimental shapes of  $\Lambda(1405)$ resonance. 
 See text for more details. 
 \label{two_exp}
 }
\vspace{-0.5cm}
\end{figure}
\end{center}

In Fig.~\ref{fig:mass} our results for the invariant mass distribution
for three different energies of the incoming $K^-$ are
compared to the experimental data. Symmetrization of the amplitudes produces a
sizable amount of background. At a kaon laboratory momentum of $p_K=581$ MeV/c
this background  distorts the $\Lambda(1405)$ shape producing cross section in
the lower part of $M_I$, while at $p_K=714$ MeV/c the strength of this
background is shifted toward the higher $M_I$ region. An ideal situation is
found for momenta around $687$ MeV/c, where the background sits below the
$\Lambda(1405)$ peak distorting its shape minimally. The peak of the resonance
shows up at $M_I^2=2.02$ GeV$^2$ which corresponds to $M_I=1420$ MeV, larger
than the nominal $\Lambda(1405)$, and in agreement with the predictions of
Ref.~\cite{Jido:2003cb} for the location of the peak when the process is
dominated by the $t_{{\bar K}N \to \pi\Sigma}$ amplitude.  The apparent width
from experiment is about $40-45$ MeV, but a precise determination would require
to remove the background mostly coming from the ``wrong'' $\pi^0 \Sigma^0$
couples due to the indistinguishability of the two pions. A theoretical analysis
permits extracting the pure resonant part by not symmetrizing the amplitude.
This is plotted in Fig. \ref{tal} as
a function of $M_I$. The width of the resonant part is $\Gamma=38$ MeV, which
is smaller than the nominal $\Lambda(1405)$ width of $50\pm 2$ MeV \cite{PDG},
obtained from the average of several experiments, and much narrower than the
apparent width of about $60$ MeV that one sees in the $\pi^- p \to K^0 \pi
\Sigma$ experiment \cite{Thomas}, which also produces a distribution peaked at
$1395$ MeV.
In order to illustrate the difference between the $\Lambda(1405)$ resonance
seen in this latter reaction and in the present one, the two
experimental distributions are compared in Fig. \ref{two_exp}. We recall
that the shape of the  $\Lambda(1405)$ in the $\pi^- p \to K^0 \pi \Sigma$ 
reaction was shown in Ref.~\cite {hyodo} to be largely built from the
 $\pi \Sigma \to \pi \Sigma$ amplitude, which is dominated by
the wider lower energy state.

\begin{center}
\begin{figure}[htb]
\vspace{-0.5cm}
   \includegraphics[width = 6.0cm]{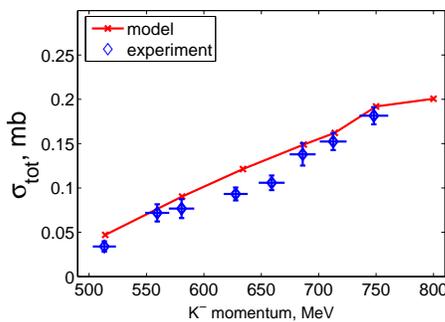}
\vspace{-0.5cm}
 \caption{Total cross section for the reaction $K^- p \to
 \pi^0 \pi^0 \Sigma^0$. Experimental data are taken from
 Ref.~\protect\cite{Prakhov}.
 \label{fig:cross}
 }
\vspace{-1.1cm}
\end{figure}
\end{center}

The invariant mass distributions shown here are not normalized, as in
experiment. But we can also compare our absolute values of
the total cross sections with those in Ref.~\cite{Prakhov}. As shown in 
Fig.~\ref{fig:cross}, our results are in excellent agreement with the data, 
in particular for the three kaon momentum values whose corresponding
invariant mass distributions have been displayed in
Fig.~\ref{fig:mass}.

In summary, we have shown, by means of a realistic model, that the  $K^- p \to
 \pi^0 \pi^0 \Sigma^0$ reaction is particularly suited to study the features
 of the second pole of the $\Lambda(1405)$ resonance, since it is largely
 dominated by a mechanism in which a $\pi^0$ is emitted prior to the $K^- p \to
 \pi^0 \Sigma^0$ amplitude, which is the one giving the largest weight to the
 second narrower state at higher enrgy.  
 Our model has proved to be accurate in reproducing both the
 invariant mass distributions and  integrated cross sections seen in 
 a recent experiment \cite{Prakhov}.  The study of the present
 reaction, complemental to the one of Ref.  \cite {hyodo} for the $\pi^- p \to
 K^0 \pi \Sigma$ reaction, has shown that the quite different shapes of the 
 $\Lambda(1405)$ resonance seen in these experiments can be interpreted in favour
 of the existence of two poles with the coresponding states having the
 characteristics predicted by the chiral theoretical calculations.  
 Besides demonstrating once more the great predictive power of the chiral
 unitary theories, this combined study of the two reactions gives the first
 clear evidence of the two-pole nature of the $\Lambda(1405)$.

{\bf Acknowledgments:}\ \ 
This work is partly supported by DGICYT contracts BFM2002-01868, BFM2003-00856,
the Generalitat de Catalunya contract SGR2001-64,
and the E.U. EURIDICE network contract HPRN-CT-2002-00311.
This research is part of the EU Integrated Infrastructure Initiative
Hadron Physics Project under contract number RII3-CT-2004-506078.


\vfill
\end{document}